\documentclass[12pt]{amsart}
\usepackage{amssymb}
\usepackage{verbatim}
\usepackage[usenames]{color}
\usepackage{hyperref}
\usepackage{enumerate}
\usepackage{rotating}

\newtheorem{thm}{Theorem}
 
\newtheorem{la}[thm]{Lemma}

\theoremstyle{definition}

\theoremstyle{remark}

\renewcommand{\phi}{\varphi}

\newcommand{\notarrow}{\kern .42em\not\kern -.42em\longrightarrow}

\newcommand{\noprint}[1]{\relax}

\title[Improved Upper Bounds for Vamos Secret Sharing Rates]{Improved Upper Bounds for the Information Rates of the Secret Sharing Schemes Induced by the Vamos Matroid}
\author{Jessica Ruth Metcalf-Burton}
\address{Mathematics Department\\
University of Michigan\\
Ann Arbor, MI 48109--1043, U.S.A.}
 \email{jmetcalf@umich.edu}

\begin{document}

\begin{abstract}
An access structure specifying the qualified sets of a secret sharing scheme must have information rate less than
or equal to one.   
The Vamos matroid induces two non-isomorphic access structures $V_1$ and $V_6$, which were shown by 
Mart\'i-Farr\'e and Padr\'o to have information rates of at least $3/4$. 
Beimel, Livne, and Padr\'o showed that the information rates of $V_1$
 and $V_6$ are bounded above by $10/11$ and $9/10$ respectively.
Here we improve those upper bounds to $19/21$ for $V_1$ and $17/19$ for $V_6$.

\end{abstract}

\maketitle

\section{Introduction}

Let $P$ be a set of participants, among whom we would like to share a secret.     
An {\it access structure} $\Gamma$ on $P$ is the collection of all 
subsets of $P$ that are {\it qualified}, i.e., allowed to reconstruct the secret.  An access structure 
$\Gamma$ is fully determined by its {\it minimal qualified subsets}, 
which are those qualified sets for which no proper subset is qualified.  Any subset of $P$ not in $\Gamma$ 
is called {\it unqualified}.  We assume that each participant in $P$ belongs to some minimal qualified subset.

We may think of the secret as belonging to a special participant called the {\it dealer}.
Intuitively, a secret sharing scheme for $\Gamma$ is a way for the dealer to select a secret and deal out
one or more {\it shares} to each participant
in such a way that qualified sets are able to reconstruct the secret by combining their shares,
while unqualified sets cannot learn any information about the secret.  

The efficiency of a secret sharing scheme can be measured in terms of its {\it information rate},
a value which indicates the size of participants' shares relative to the size of the secret.
The information rate will always be between zero and one \cite{Csirmaz}.  
Mart\'i-Farr\'e and Padr\'o \cite{MFP} showed that any
access structure with information rate greater than $\frac{2}{3}$ is induced by a matroid.  

It is not yet known how to determine the information rates of the access structures induced by a particular matroid.
One matroid currently under consideration is the Vamos matroid, 
which induces two non-isomorphic access structures $V_1$ and $V_6$.
Each of these access structures is known to have an information rate of at least $3/4$ \cite{MFP}, and
Beimel, Livne, and Padr\'o showed that the information rates of $V_1$
 and $V_6$ have upper bounds of $10/11$ and $9/10$ respectively \cite{BLP}
(Beimel et. al. refer to $V_8$ rather than $V_1$, but the two are isomorphic and 
$V_1$ is notationally more convenient for our purposes). 
Here we improve those upper bounds to $19/21$ for $V_1$ and $17/19$ for $V_6$.

\section{Secret Sharing Schemes}

We now give a more precise definition of a secret sharing scheme, following the ideas of 
Csirmaz \cite{Csirmaz} and Mart\'{i}-Farr\'{e} and Padr\'{o} \cite{MFP}. 
Let $\Sigma$ be a collection of random variables consisting of one random variable $S$ for the secret and,
for each participant $x \in P$, a random variable for the share belonging to $x$.

For any participant $x \in P$, we use $H(x)$ to denote the Shannon entropy of the corresponding 
random variable, and for any nonempty
subset $X \subseteq P \cup \{S\}$, we use $H(X)$ to denote the joint entropy of the 
random variables for all elements of $X$.  We use $H(X|Y)$ to denote conditional entropy for nonempty sets 
$X,Y \subseteq P \cup \{S\}$.  Recall that by definition $H(X|Y)=H(X \cup Y)-H(Y)$.

We call $\Sigma$ a {\it (perfect) secret sharing scheme for $\Gamma$} if it has the following properties:
\begin{itemize}
\item If $X \in \Gamma$ then  $H(S|X)=0$, that is, the participants in $X$ are able to combine their shares to 
completely determine the value of the secret.
\item If $X \notin \Gamma$ then $H(S|X)=H(S)$, that is, the uncertainty about the secret does not change even when all participants in $X$ pool their shares.
\end{itemize}


Given a secret sharing scheme $\Sigma$ and a participant $x \in P$, the {\it information rate of $x$} is defined by
\begin{equation*} \rho(x)=\frac{H(S)}{H(x)} .\end{equation*}
The information rate of $\Sigma$, $\rho(\Sigma)$, is the minimum information rate over all participants in $P$.
For an access structure $\Gamma$, the information rate $\rho(\Gamma)$ is the supremum of $\rho(\Sigma)$ over all
$\Sigma$ that are secret sharing schemes for $\Gamma$.

\section{Normalized Entropy}

Fix a (perfect) secret sharing scheme $\Sigma$.  
We define the {\it normalized entropy} of a nonempty set $X \subseteq P$ by
$$h(X)=\frac{H(X)}{H(S)}$$
and the {\it conditional normalized entropy of $X$ given $Y$} for nonempty sets $X,Y \subseteq P$ by
$$h(X|Y)=\frac{H(X|Y)}{H(S)}.$$
The entropy function $H$ is nonnegative.
We may assume that $H(S)$ is strictly positive, because if $H(S)=0$ then $H(S|X)=0$ for every $X \subseteq P$, meaning that every 
set of participants is qualified and there is nothing left to investigate.  Thus the normalized entropy and conditional normalized entropy are
well-defined.  We observe that 
the information rate for a participant $x \in P$ is the reciprocal of the normalized entropy for that participant:
\begin{equation}\label{h_and_rho}\rho(x)=\frac{1}{h(x)}.\end{equation}

The normalized entropy is monotone and submodular, as can be shown by dividing through the appropriate 
inequalities for the entropy function by the positive quantity $H(S)$.
Some additional useful facts about $h$ 
are described in the following lemmas.  We assume that $X,Y$ are nonempty subsets of $P$. We will frequently omit the
symbol for set union, writing $XY$ for $X \cup Y$.

\begin{la}\label{addSqual} If $X \in \Gamma$ then $h(X)=h(XS)$.
\end{la}
\begin{proof}
From the definitions of (perfect) secret sharing scheme and conditional entropy, if $X \in \Gamma$ then
$$0 = H(S|X) = H(XS)-H(X).$$  
Dividing through by $H(S)$ and rearranging gives the desired result.
\end{proof}

\begin{la}\label{addSunqual} If $X \notin \Gamma$ then $1 = h(XS)-h(X)$.
\end{la}
\begin{proof}
From the definitions of (perfect) secret sharing scheme and conditional entropy, if $X \notin \Gamma$ then
\begin{equation*} H(S)=H(S|X)=H(XS)-H(X). \end{equation*}
Dividing through by $H(S)$ gives the desired result.
\end{proof}

\begin{la}\label{add1unqual} If $X \notin \Gamma$ but $XY \in \Gamma$ then $1 \leq h(XY) -h(X)$.
\end{la}
\begin{proof}
Using the monotonicity of $h$ and lemmas \ref{addSqual} and \ref{addSunqual},
\begin{equation*}1 = h(XS)-h(X) \leq h(XYS) - h(X) = h(XY)-h(X).\end{equation*}    
\end{proof}

Lemma \ref{add1unqual} says that if $X$ is unqualified and adding the participants in $Y$ produces a qualified
set, then the participants in $Y$ must contribute at least 1 to the normalized entropy of $X$.
A slight reformulation of this is the following lemma, which says that 
if adding a participant $r$ to an unqualified superset of $X$ produces a qualified set, 
then $r$ must contribute at least 1 to the 
normalized entropy of $X$.

\begin{la}\label{supersetunqual}  If $XY \notin \Gamma$ but $XY \cup \{r\} \in \Gamma$ then
$1 \leq h(X \cup \{r\}) - h(X).$
\end{la}
\begin{proof}
By lemma \ref{add1unqual} 
\begin{eqnarray*}
1 &\leq& h(XY \cup \{r\})-h(XY)
\end{eqnarray*}
and by the submodularity of $h$
\begin{eqnarray*}
h(X) + h(XY \cup \{r\}) &\leq& h(XY) + h(X \cup \{r\}).
\end{eqnarray*}
If we add these inequalities, cancel terms, and rearrange, we get the desired result.
\end{proof}

\begin{la}\label{qual2ways} If $X\cap Y \notin \Gamma$ but $X, Y \in \Gamma$ then
\begin{eqnarray*}
h(X\cap Y) + h(XY) + 1 &\leq& h(X) + h(Y). \end{eqnarray*}  
\end{la}
\begin{proof}
By the submodularity of $h$,
\begin{eqnarray*}h((X\cap Y)S) + h(XYS) &\leq& h(XS) + h(YS).\end{eqnarray*}  
Since $X, Y, XY \in \Gamma$, adding $S$ to any of these sets
does not change their normalized entropy.  However, by lemma \ref{addSunqual} 
$$h((X\cap Y)S) = h(X\cap Y)+1.$$
\end{proof}

\section{Matroids and Secret Sharing Schemes}

A {\it matroid} $M$ over a finite set $Q$ is a collection $\mathfrak{I}$, called the {\it independent}
subsets of $Q$, such that 
\begin{itemize}
\item the empty set is independent,
\item subsets of independent sets are independent, and
\item if $X,Y$ are independent with $|X|=|Y|+1$ there is $x \in X\setminus Y$ such that $Y \cup \{x\}$ is independent.
\end{itemize} 

Any set that is not independent is {\it dependent}.  Maximal independent sets are called {\it bases}, and
minimal dependent sets are called {\it circuits}.  A matroid may also be specified in terms of its bases or
circuits.   For a more thorough introduction to matroids we refer the reader to \cite{Welsh}. 

Given a matroid $M$ over $Q$, each element $x \in Q$ induces an access structure $\Gamma_x$ over the participants 
$Q \setminus \{x\}$.
The minimal qualified sets of $\Gamma_x$ are those subsets $Y \subseteq Q\setminus\{x\}$ for which 
$Y \cup \{x\}$ is a circuit in the matroid $M$.  
Intuitively, if $Y \cup \{x\}$ is a circuit then the value of $x$ can be determined from the elements of $Y$.
More discussion of matroids and access structures may be found in \cite{MFP}.

\section{The Vamos Matroid}

We define the Vamos matroid on the set $\{v_1, \dots, v_8\}$ as follows.
First define the {\it Vamos pairs} $A$, $B$, $C$, and $D$ by
 $A=\{v_1, v_2\}$, $B=\{v_3, v_4\}$, $C=\{v_5, v_6\}$, and $D=\{v_7, v_8\}$.  
The Vamos matroid on $ABCD$ is the matroid 
whose independent sets are all sets of size less than 5 except for the sets $AB$, $AC$, $BC$, $BD,$ and $CD$.
Thus the sets $AB$, $AC$, $BC$, $BD$, and $CD$ are circuits in the Vamos matroid.  Any set of fewer than 4 elements is
independent, and any set with more than four elements is dependent.

In the following discussion when we speak about circuits, independent sets, and dependent sets, 
we mean these terms with respect to the Vamos matroid.

Because of symmetries there are, up to isomorphism, two access structures induced by the Vamos matroid.  
One is the structure $V_1$, where $v_1$ is the dealer.  The other is $V_6$, where $v_6$ is the dealer.
For convenience we shall consider each of $V_1$ and $V_6$ to be an access structure on eight participants, thinking of the dealer
as a participant who is individually qualified to recover the secret.  Recall that the other minimal qualified sets will be
those sets of participants who, with the inclusion of the dealer, form a circuit in the Vamos matroid. 
Note that this means any qualified set which does not contain the dealer must include at least 3 participants. 

As in \cite{BLP}, for a fixed secret sharing scheme $\Sigma$ on $V_1$ or $V_6$ we define
\begin{equation*}
\lambda = \left( \displaystyle\max_{1 \leq i \leq 8}h(P_i) \right) - 1
\end{equation*}
so that for each participant 
\begin{equation}\label{lambda}
h(v_i) \leq 1 + \lambda.
\end{equation}
We note that by equation \eqref{h_and_rho} the information rate of the scheme will then be 
\begin{equation}\label{schemeinfo}
\rho(\Sigma)=\min_{1 \leq i \leq 8}\frac{1}{h(P_i)} = \frac{1}{1+\lambda}.
\end{equation}

\begin{la}\label{upbounds} Let $X, Y$ be distinct Vamos pairs with $XY$ a circuit.  If the dealer is a member of $Y$, then
\begin{enumerate}[(i)]
\item $h(Y|X) \leq 1 + \lambda$
\item $h(X|Y) \leq 1 + 2\lambda.$
\end{enumerate}
\end{la}

\begin{proof}
Let $Y=\{s,t\}$ where $s$ is the dealer, and let $X=\{p,q\}$.
\begin{enumerate}[(i)]
\item Since $XY$ is a circuit containing the dealer, $X\cup\{t\}$ is a qualified set.  Thus by lemma \ref{addSqual},
the submodularity of $h$, and equation \eqref{lambda},
\begin{equation*}
h(Y|X)=h(\{t\}|X) \leq h(\{t\}) \leq 1 + \lambda.
\end{equation*}
\item The set $\{p,t\}$ is unqualified, as it is a set of size 2 that does not include the dealer.  
The set $X \cup \{t\}$ is qualified.  Thus by lemma~\ref{qual2ways},
\begin{equation*} h(\{p,t\})+h(XY)+1 \leq h(\{p\} \cup Y) + h(X \cup \{t\}). \end{equation*}
Subtracting $h(Y)$ from both sides, rearranging, and using equation \eqref{lambda} gives us
\begin{eqnarray*}
h(X|Y) &\leq& h(\{p\}|Y) + h(\{q\}|\{p,t\}) -1 \\
&\leq& h(\{p\}) + h(\{q\}) -1\\
&\leq& 2(1+\lambda) - 1\\
&=&1 + 2\lambda.
\end{eqnarray*} 
\end{enumerate}
\end{proof}

\begin{la}\label{3lambda}Let $X,Y$ be distinct Vamos pairs with $XY$ a circuit.
If neither $X$ nor $Y$ contains the dealer, then 
\begin{equation*}h(Y|X) \leq 1 + 3\lambda.\end{equation*}
\end{la}
\begin{proof}
Let $Y=\{p,q\}$.  Take $r$ to be one of the two participants that is neither in $XY$ nor in the Vamos pair of
the dealer.  Then we will have $X \cup \{r\} \notin \Gamma$, since adding the dealer to these three elements does not 
produce a circuit.  
We will have $X \cup \{p,r\}, X \cup \{q,r\} \in \Gamma$, since each of these is an independent set with four 
participants.  Then by lemma~\ref{qual2ways} we have
\begin{eqnarray*}
h(X \cup \{r\}) + h(XY \cup \{r\}) +1 &\leq& h(X \cup \{p,r\}) + h(X \cup \{q,r\}).
\end{eqnarray*}
Since $XY \notin \Gamma$, by lemma \ref{add1unqual} we have
\begin{eqnarray*}
h(XY) + 1 &\leq& h(XY \cup \{r\}).
\end{eqnarray*}
We get the following from the submodularity of $h$:
\begin{eqnarray*}
h(X \cup \{p\}) &\leq& h(X) + h(\{p\})\\
h(X \cup \{p,r\}) &\leq& h(X \cup \{p\}) + h(\{r\})\\
h(X \cup \{q,r\}) &\leq& h(X \cup \{r\}) + h(\{q\}).
\end{eqnarray*}
Finally, from equation \eqref{lambda},
\begin{eqnarray*}
h(\{p\}) &\leq& 1 + \lambda\\
h(\{q\}) &\leq& 1 + \lambda\\
h(\{r\}) &\leq& 1 + \lambda.
\end{eqnarray*}
Adding the inequalities above, canceling terms, and writing as conditional entropy gives us the bound specified.
\end{proof}

\begin{la}\label{at_least_2}Let $X,Y$ be distinct Vamos pairs with $XY$ independent.  If the dealer is not a member of $X$, then
\begin{equation*} 2 \leq h(Y|X).  \end{equation*}
\end{la}
\begin{proof}

{\it Case 1:} Assume that the dealer is not a member of $Y$.

Let $Y=\{p,q\}$.  Since $XY$ is qualified but $X \cup \{p\}, X \cup \{q\}$ are not, by lemma \ref{add1unqual} we get
the inequalities
\begin{align*}
1 \leq& h(XY) - h(X \cup \{p\})\\
1 \leq& h(XY) - h(X \cup \{q\}).
\end{align*}
By the submodularity of $h$, 
$$ h(XY) + h(X) \leq h(X \cup \{p\}) + h(X \cup \{q\}).$$
Adding the above three inequalities, canceling terms, and rearranging gives the desired result.

{\it Case 2:} Assume that $Y=\{s,t\}$ where $s$ is the dealer.
Since $XY$ is independent, $X \cup \{t\}$ is unqualified.  Thus by lemma \ref{addSunqual}
\begin{equation*}
1 \leq h(XY) - h(X \cup \{t\}).
\end{equation*}
Let $r$ be any participant not in $XY$.  Then $X \cup \{r\}$ is also unqualified.  Since $X \cup \{t,r\}$ is 
qualified, lemma \ref{supersetunqual} tells us that
\begin{equation*}
1 \leq  h(X \cup \{t\}) -h(X).
\end{equation*}
Adding the above two inequalities gives the desired result.
\end{proof}

Although the previous lemma is stated in general terms, it will only apply to the Vamos pairs $A$ and $D$, as any other two
distinct Vamos pairs are dependent. 

\section{New Bounds For Lambda}

In \cite{BLP} Beimel, Livne, and Padr\'o found by looking at the Zhang-Yeung non-Shannon inequality \cite{ZY}
that when $v_6$ is the dealer, $1/9 \leq \lambda$, and when $v_1$ is the dealer,
$1/10 \leq \lambda$.  Here we improve those bounds by looking at other non-Shannon inequalities from \cite{Dougherty}.

\begin{thm}\label{C}
If the dealer is a member of $C$, then $2/17 \leq \lambda.$
\end{thm}
\begin{proof}
We use Dougherty, Freiling, and Zeger's inequality (i) from \cite{Dougherty}, which may be written
\begin{align*}\tag{DFZi}\label{Di}
0 \leq& - 3h(A) - 5h(B)- 3h(C) + 8h(AB) \\
& + 6h(AC) - 2h(AD) + 6h(BC)+ 2h(BD)\\
& + 2h(CD)- 9h(ABC)- 2h(BCD).
\end{align*}
Since $A, AB, BD \notin \Gamma$ and $ABC, BCD, AD \in \Gamma$, 
from lemmas \ref{add1unqual}, \ref{supersetunqual}, and \ref{at_least_2} we obtain the following inequalities, which we add 
to \eqref{Di} with the indicated multiplicities:
\begin{eqnarray*}
9[1 &\leq& h(ABC)-h(AB)]\\
2[1 &\leq& h(BCD)-h(BD)]\\
2[2 &\leq& h(AD)-h(A)]\\
1 &\leq& h(AB)-h(A).
\end{eqnarray*}
After canceling terms, the sum of inequalities yields
$$16 \leq -6h(A) -5h(B) -3h(C) + 6h(AC) +6h(BC) +2h(CD).$$
Rearranging, we obtain
\begin{eqnarray*}
16 &\leq& 6[h(AC)-h(A)] + 5[h(BC)-h(B)] \\
&& + [h(BC)-h(C)] + 2[h(CD)-h(C)]
\end{eqnarray*}
which may be further rewritten as 
$$16 \leq 6h(C|A) + 5h(C|B) +h(B|C) +2h(D|C).$$
Replacing each conditional normalized entropy by its upper bound from lemma \ref{upbounds}, 
we get 
$$16 \leq 6(1+\lambda) + 5(1+\lambda) + (1 + 2\lambda) + 2(1 + 2\lambda).$$
This simplifies to
$$16 \leq 14 + 17\lambda$$
and we conclude that $2/17 \leq \lambda.$
\end{proof}

\begin{thm}\label{A}
If the dealer is a member of $A$, then $2/19 \leq \lambda.$
\end{thm}
\begin{proof}
We begin with inequality (iv) from \cite{Dougherty}, which is
\begin{align*}\tag{DFZiv}\label{Div}
0 \leq& -h(A) -5h(B) -5h(C) +6h(AB)\\
& + 6h(AC) - 2h(AD) +8h(BC) +2h(BD)\\
& + 2h(CD) -9h(ABC)-2h(BCD).
\end{align*}

To this we add four inequalities (with indicated multiplicities) to cancel out the terms with $h(AD),$ $h(ABC),$ and $h(BCD)$.

Since $AD \in \Gamma$ but $D \notin \Gamma$, by lemma \ref{at_least_2} 
\begin{eqnarray*}2[2 &\leq& h(AD) - h(D)].\end{eqnarray*}

Since $ABC, BCD \in \Gamma$ and $BC, CD \notin \Gamma$, by lemma \ref{add1unqual} 
\begin{align*}9[1 \leq& h(ABC)-h(BC)]\\
2[1 \leq& h(BCD) - h(CD)].\end{align*}

Finally, since $C$ combined with either participant in $D$ will still be an 
unqualified set, by lemma \ref{supersetunqual} we have
\begin{eqnarray*}1 &\leq& h(BC)-h(C).\end{eqnarray*}

After adding the above inequalities to \eqref{Div}, simplifying, and canceling terms we are left with 
\begin{eqnarray*}
16 &\leq& -h(A)-5h(B) - 6h(C) - 2h(D) + 6h(AB)\\
&& +  6h(AC) + 2h(BD) 
\end{eqnarray*}
which can be rearranged into
\begin{eqnarray*}
16 &\leq& 5h(A|B) + h(B|A) + 6h(A|C) + 2h(B|D).
\end{eqnarray*}
Using the bounds found in lemmas \ref{upbounds} and \ref{3lambda},
we have
\begin{eqnarray*}
16 &\leq& 5(1+\lambda) + (1+2\lambda) + 6(1+\lambda) + 2(1+3\lambda) 
\end{eqnarray*}
and we conclude that $2/19 \leq \lambda$.

\end{proof}

The method of canceling terms used here was generalized and 
applied to the other inequalities in \cite{Dougherty}, after appropriate permutations of letters in the
other inequalities. However, only bounds for $\lambda$ weaker than those shown here were obtained.

\section{Conclusion}

By theorem \ref{C} and equation \eqref{schemeinfo}, for any secret sharing scheme $\Sigma$ on $V_6$ we have
\begin{equation*}\rho(\Sigma)=\frac{1}{1+\lambda} \leq \frac{17}{19}\end{equation*}
and thus by the definition of information rate for an access structure, 
\begin{equation*}
\rho(V_6) \leq \frac{17}{19}.
\end{equation*}
Similarly, by theorem \ref{A} and equation \eqref{schemeinfo},
\begin{equation*}
\rho(V_1) \leq \frac{19}{21}.
\end{equation*}
These are improvements to the best previously known upper bounds for the information rates of the access structures induced
by the Vamos matroid.

\section{Acknowledgments}
The author would like to thank Andreas Blass.


\begin{thebibliography}{6}



\bibitem{BLP}
A. Beimel, N. Livne, C. Padr\'{o}.  ``Matroids Can Be Far From Ideal Secret Sharing''.  {\it Lecture Notes in Computer
Science}, vol.~4948, pp. 194-212, 2008.

\bibitem{Csirmaz}
L. Csirmaz.  ``The Size of a Share Must be Large.'' {\it Journal of Cryptology}, vol.~10, n. 4, pp. 223-231, 1997.

\bibitem{Dougherty}
R. Dougherty, C. Freiling, K. Zeger.  ``Six New Non-Shannon Information Inequalities''.  {\it 2006 IEEE International
Symposium on Information Theory}, pp.233-236, 2006.

\bibitem{MFP}
 J. Mart\'{i}-Farr\'{e}, C. Padr\'{o}.  ``On Secret Sharing Schemes, Matroids and Polymatroids.''  Cryptology ePrint Archive, Report 2006/077, 2006.\\
  \url{http://eprint.iacr.org/}.




\bibitem{Welsh}
D. J. A. Welsh.  {\it Matroid Theory}.  London:  Academic Press, 1976.  

\bibitem{ZY}
Z. Zhang, R. W. Yeung. ``On Characterization of Entropy Function via Information Inequalities.'' {\it IEEE Transactions
on Information Theory}, vol.~44, n.~4, pp. 1440-1452, 1998.

\end{thebibliography}
\end{document}